\documentclass[12pt,a4]{article}
\usepackage{epsf}  
\usepackage{cite}  
\begin{document}
%
\begin{center}
{\LARGE\sf Cascade in Muonic and Pionic Atoms\\[2mm] with $Z=1$}\\[5mm]
{\sc V.~E.~Markushin}\\[2mm] 
{\it Paul Scherrer Institute, CH-5232 Villigen, Switzerland}\\[5mm]
{December 2, 1998}\\[5mm]
{\small Invited talk at EXAT-98, Ascona, July 19--24, 1998 \\ 
        (to be published in Hyperfine Interactions)}\\[5mm]
\end{center} 

\begin{abstract}
   Recent theoretical and experimental studies of the exotic atoms with
$Z=1$ are reviewed.  An interplay between the atomic internal and
external degrees of freedom is essential for a good description of the
atomic cascade.
   The perspective of {\it ab initio} cascade calculations is outlined.
\end{abstract}

%
%
%

\section{Introduction}

   When heavy negative particles ($\mu^-$, $\pi^-$, $\bar{p}$, etc.)
stop in matter, they usually form exotic atoms in highly excited states
with principal quantum number $n\sim\sqrt{m/m_e}$ where $m$ is the
reduced mass of the exotic atom and $m_e$ is the electron mass.
   The exotic-atom formation is followed by an atomic cascade consisting
of multistep transitions to lower atomic states.
   For hadronic atoms, 
the atomic cascade is a complete life
history because the hadrons get absorbed by the nuclei with high
probability before reaching the ground state.
   Muonic atoms (where the absorption is weak) de-excite to the ground
state and engage in various reactions (muon catalyzed fusion, muon
transfer, molecular formation) with initial conditions determined by the
atomic cascade.
   In both cases, the atomic cascade can reveal important information
about the properties of exotic atoms and reactions with atoms in excited
states.
   This paper, supplementing the earlier reviews
\cite{LB62,Ma90,Ko92,Ma94}, focuses on the recent progress in
theoretical studies of the atomic cascade in light muonic and pionic
atoms.

\section{\label{SecCM} Cascade Mechanisms}

   A brief summary of the cascade processes in the exotic atoms with
$Z=1$ is given in Table~\ref{TabCM}.  The radiative de-excitation and
the nuclear absorption (in hadronic atoms) do not depend on experimental
conditions directly.  All other processes occur in collisions with
surrounding atoms and their rates are proportional to the
hydrogen density and usually depend on energy.

   At least tree cascade mechanisms are essential for the basic
understanding of the atomic cascade \cite{LB62}: the radiative
transitions, the external Auger effect, and the Stark mixing.  In this
paper, the cascade models, which include these three mechanisms only,
will be called the minimal cascade model\footnote{In the literature, it
has also been called the standard cascade model (SCM).} (MCM).

\begin{table}
\begin{center}
\begin{small}
\renewcommand{\arraystretch}{1.2}
\begin{tabular}{|l|c|c|c|} \hline \hline
 Mechanism & Example & $E$-dependence & Refs. \\
\hline
 Radiative  &
 $(\mu p)_i \to (\mu p)_f + \gamma$  &
 none  &
 see \cite{LB62} \\
\hline
 External Auger effect &
 $(\mu p)_i +{\rm H}_2 \to (\mu p)_f + e^- +{\rm H}_2^+$  &
 weak &
 \cite{LB62,BuPo} \\
\hline
 Stark mixing &
 $(\mu p)_{nl} + {\rm H} \to (\mu p)_{nl^\prime} + {\rm H}$ &
 moderate &
 \cite{LB62,Ve,BL80,Ma81,PP96,TH97,PP98} \\
\hline
 Elastic scattering &
 $(\mu p)_{n} + {\rm H} \to (\mu p)_{n} + {\rm H}$ &
 strong &
 \cite{PP98,MP86,Cz96,By96,PPG98} \\
\hline
 Coulomb transitions &
 $(\mu p)_{n_i} + p \to (\mu p)_{n_f} + p$, $n_f < n_i$   &
 strong &
 \cite{BF,Me88,Cz90,Cz94CT,PS96,So98,Kr98} \\
 \hline
 Transfer (isotope exchange) &
 $(\mu p)_{n} + d \to (\mu d)_{n} + p$ &
 strong &
 \cite{MP84,Kr88,GPS93,Cz94T} \\
\hline
 Absorption &
 $(\pi^-p)_{nS} \to \pi^0+n,\ \gamma+n$ &
 none &
 see \cite{LB62} \\
\hline \hline
\end{tabular}
\caption{\label{TabCM}%
Cascade processes in exotic atoms with $Z=1$ and their energy
dependence.
}
\end{small}
\end{center}
\end{table}

\begin{figure}
\mbox{\hspace{20mm} (a) \hspace{60mm} (b)}\\[-\baselineskip]
\mbox{%
\mbox{\epsfysize=6cm\epsffile{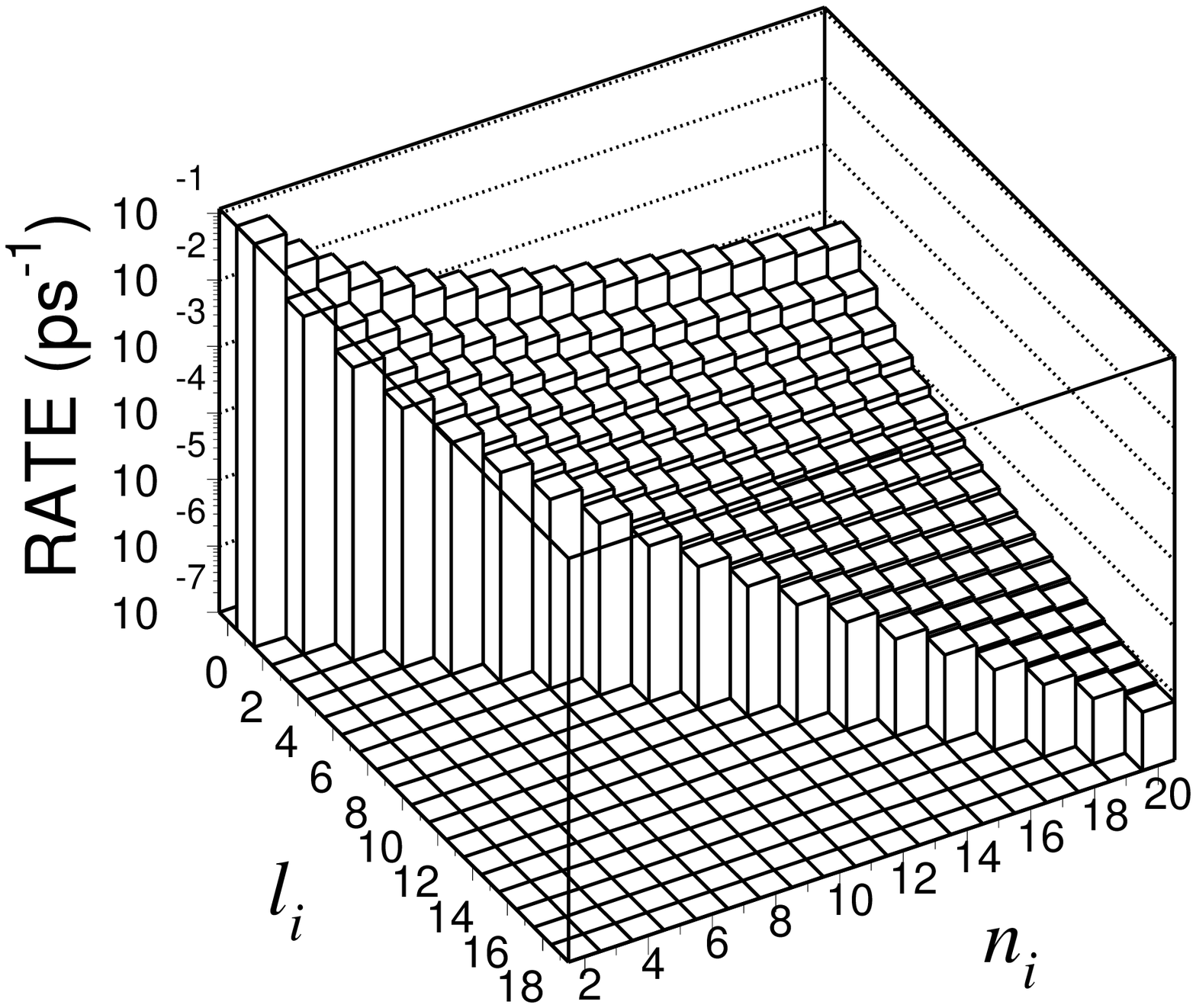}}%
\mbox{\epsfysize=6cm\epsffile{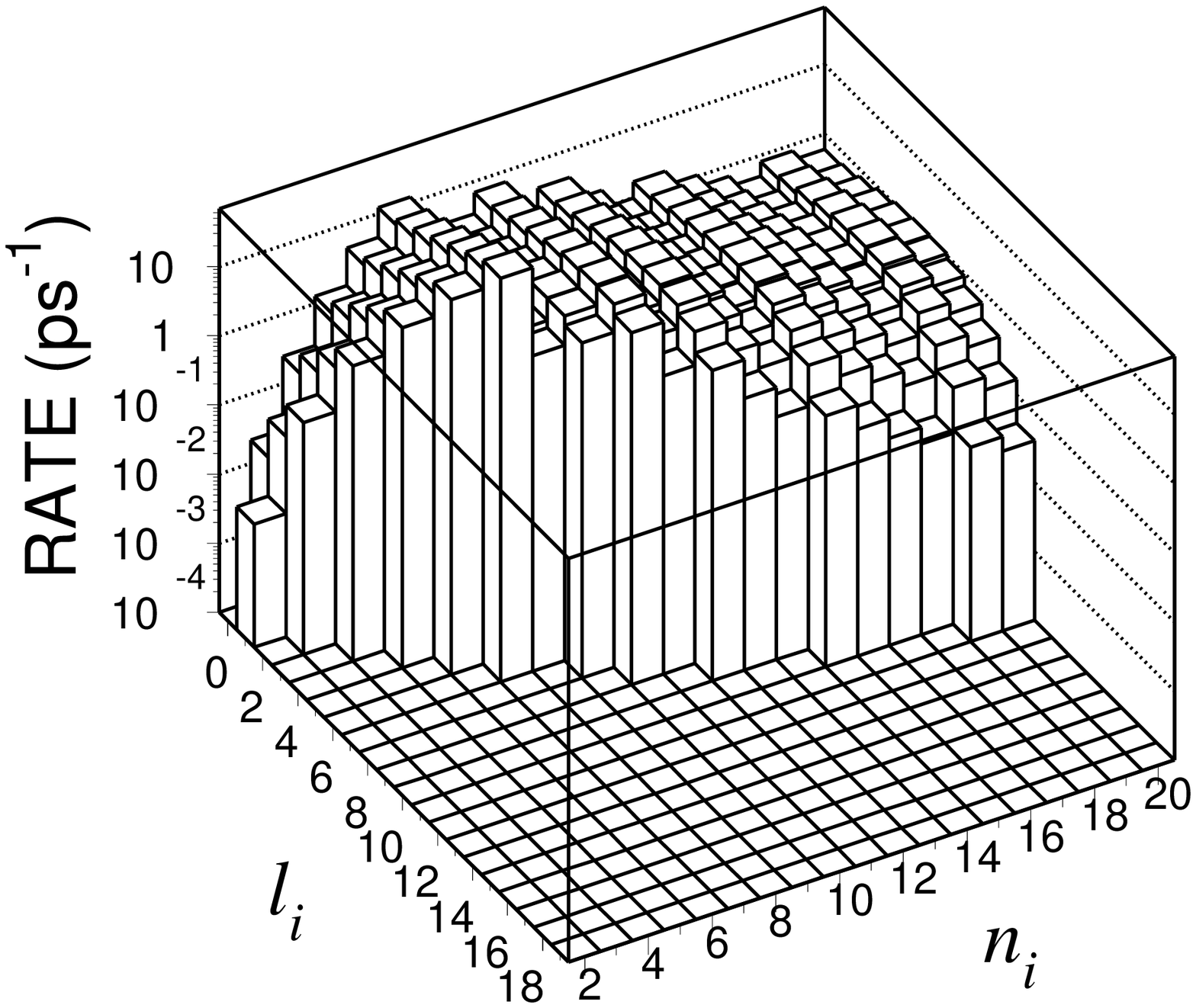}}
}
\caption{\label{FigRadAug}%
The rates of (a) radiative and (b) Auger de-excitation (LHD) in
muonic hydrogen.}
\end{figure}

   Figure~\ref{FigRadAug} demonstrates the $nl$-dependence of the total
radiative and Auger de-excitation rates for muonic hydrogen.  The main
features of these de-excitation mechanisms were discussed in
\cite{LB62,Ma90}.

   The Auger rates calculated in the Born approximation
(Fig.~\ref{FigRadAug}b) are energy independent.  The eikonal
approximation \cite{BuPo} predicts a rather weak energy dependence, with
the results being very close to the ones in the Born approximation for
$n\leq 6$ and for a kinetic energy of the order of 1~eV.
The initial and final state interactions in the Auger transitions were
discussed in \cite{Me88}, however, no detailed calculations have been
done.


   The Stark mixing corresponds to transitions among the $nl$-sublevels
with the same $n$.  It is a very fast collisional process because the
exotic atoms with $Z=1$ are small and electroneutral and have no
electrons, so that they can easily pass through the regions of the
strong electric field inside ordinary atoms.  When the Stark mixing rate
is much larger than all other transition rates, the statistical
population of the $nl$-sublevels is determined by the principle of
detailed balance.

\begin{figure}
\mbox{\hspace{20mm} (a) \hspace{60mm} (b)}\\[-\baselineskip]
\mbox{%
\mbox{\epsfysize=6cm\epsffile{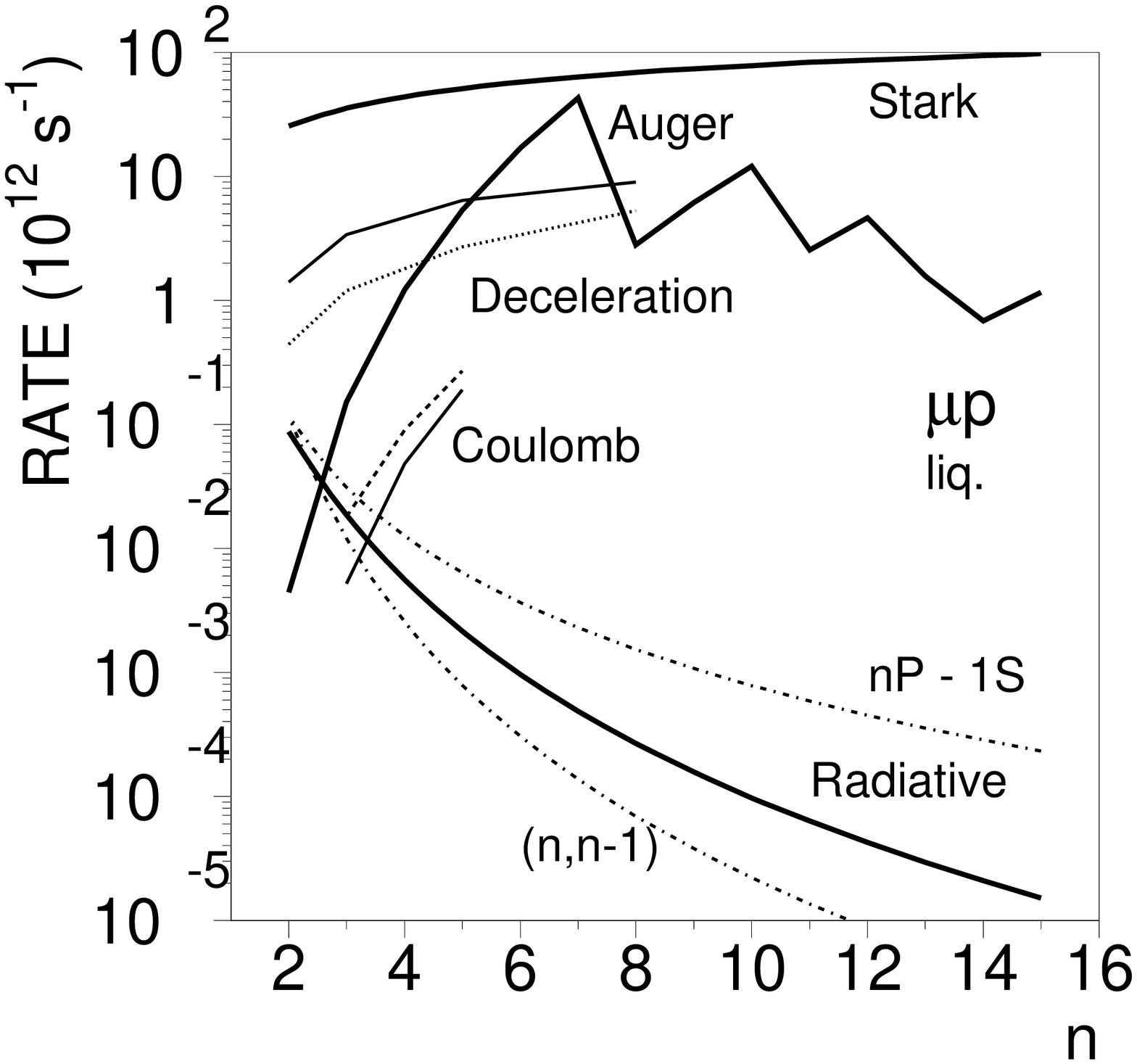}}%
\mbox{\epsfysize=6cm\epsffile{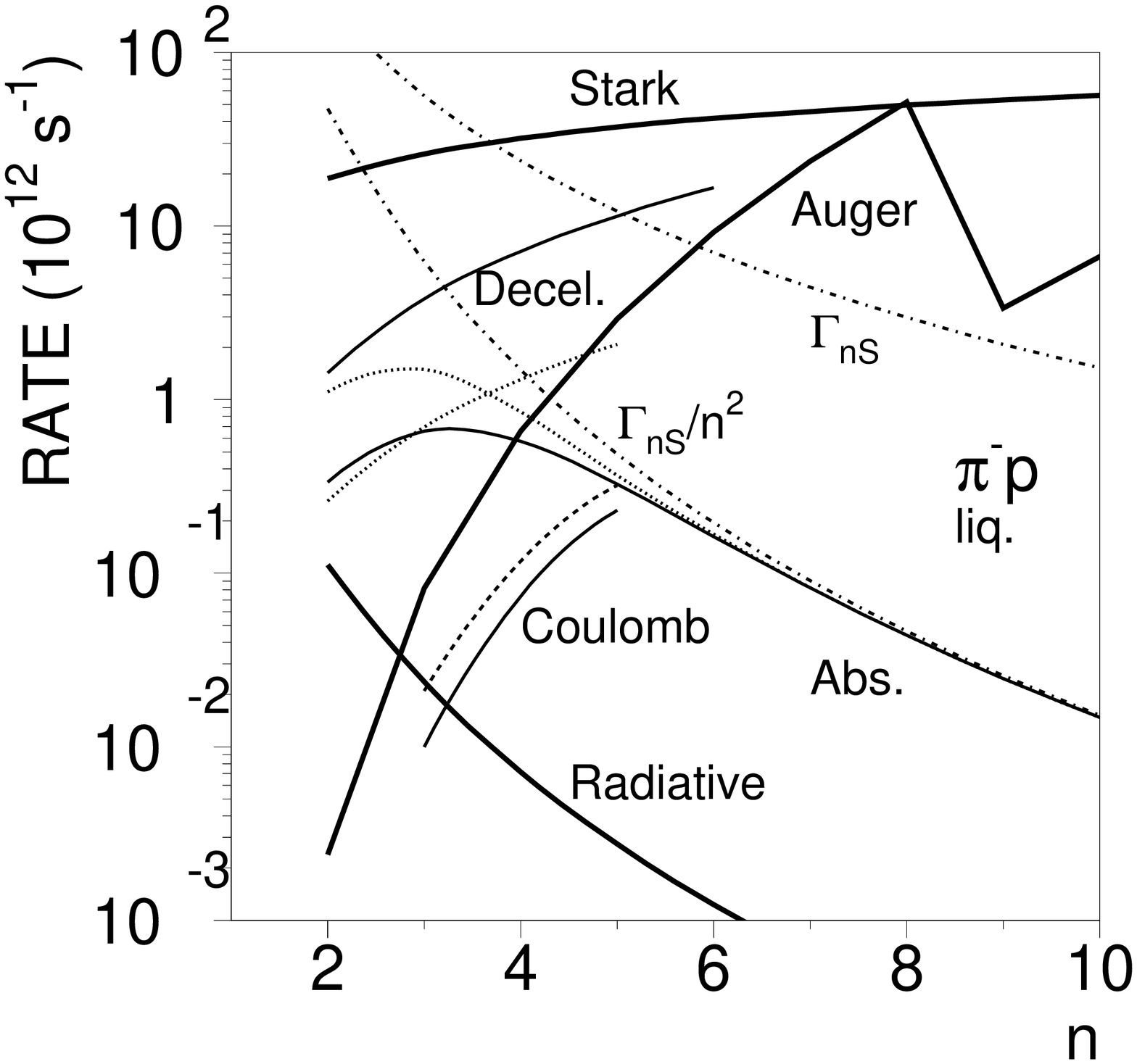}}
}
\caption{\label{Figmuppip}%
The effective (statistical average) rates for
(a) muonic hydrogen \cite{Ma94} and (b) pionic hydrogen \cite{As95}
in liquid hydrogen.
The {\it effective} absorption rates for the $\pi^-p$ defined
as in \cite{LB62} and the deceleration rates correspond to the kinetic energy
$E=2\;$eV (solid line) and $E=30\;$eV (dotted line).
The Coulomb de-excitation rates
for $E=2\;$eV (solid line) and $E=0.04\;$eV (dashed line) are from
\cite{PS96,So98}.
}
\end{figure}

   The relative importance of the various cascade processes in muonic
and pionic hydrogen is demonstrated in Figs.~\ref{Figmuppip}a,b.
   The cascade models of the exotic atoms with $Z=1$ are listed in
Table~\ref{TabMod}; they are of two types.  In one group, there are
various implementations of the MCM \cite{LB62,BL80,Ma81,TH97}
where the kinetic energy is assumed to be constant through the whole
cascade.
   The other group \cite{Ma94,As95,Ru95,AM97} consists of detailed
kinetics models which take the energy evolution during cascade into
account by explicit treatment of acceleration and deceleration
mechanisms.

\begin{table}
\begin{small} 
\begin{tabular}{|l|l|lll|}
\hline
Model & Systems  &  Coulomb & Elastic & $E$-evolution \\
\hline
Leon, Bethe (1962)
 \cite{LB62} & $\pi^-p$, $K^-p$
             & $-$   & $-$   & $-$  \\
Borie, Leon (1980)
 \cite{BL80} & $\mu p$, $\pi^-p$,  $K^-p$, $\bar{p}p$
             & $-$   & $-$   & $-$ \\
Markushin (1981)
 \cite{Ma81} & $\mu p$, $\mu d$
             & $-$   & $-$   & $-$ \\
Reinfenr\"oter et al. (1988)
 \cite{Re88} & $\bar{p}p$            
             & $-$   & $-$   & $-$ \\
Czaplinski et al. (1990)
 \cite{Cz90Casc} & $\mu p$, $\mu d$
             & $-$   & $-$   & $-$ \\
Markushin (1994)
 \cite{Ma94} & $\mu p$, $\mu d$
             & $+$   & $+$   & $+$ \\
Czaplinski et al. (1994)
 \cite{Cz94T} & $\mu p$, $\mu d$
             & $+$   & $-$   & $-$ \\
Aschenauer et al. (1995)
 \cite{As95,Ru95} & $\pi^-p$
             & $+$   & $+$   & $+$ \\
Aschenauer, Markushin (1997)
 \cite{AM97} & $\mu p$, $\mu d$
             & $+$   & $+$   & $+$ \\
Terada, Hayano (1997)
 \cite{TH97} & $\pi^-p$, $K^-p$, $\bar{p}p$
             & $-$   & $-$   & $-$ \\
\hline
\end{tabular}
\caption{\label{TabMod}%
The cascade processes included in various models of the lightest
exotic atoms in addition to the MCM containing the radiative and Auger
de-excitation and the Stark mixing.
}
\end{small}
\end{table}

\section{X-Ray Yields}

\begin{figure}
\mbox{\hspace{25mm} (a) \hspace{60mm} (b)}\\[-0.5\baselineskip]
\mbox{%
\mbox{\epsfysize=65mm\epsffile{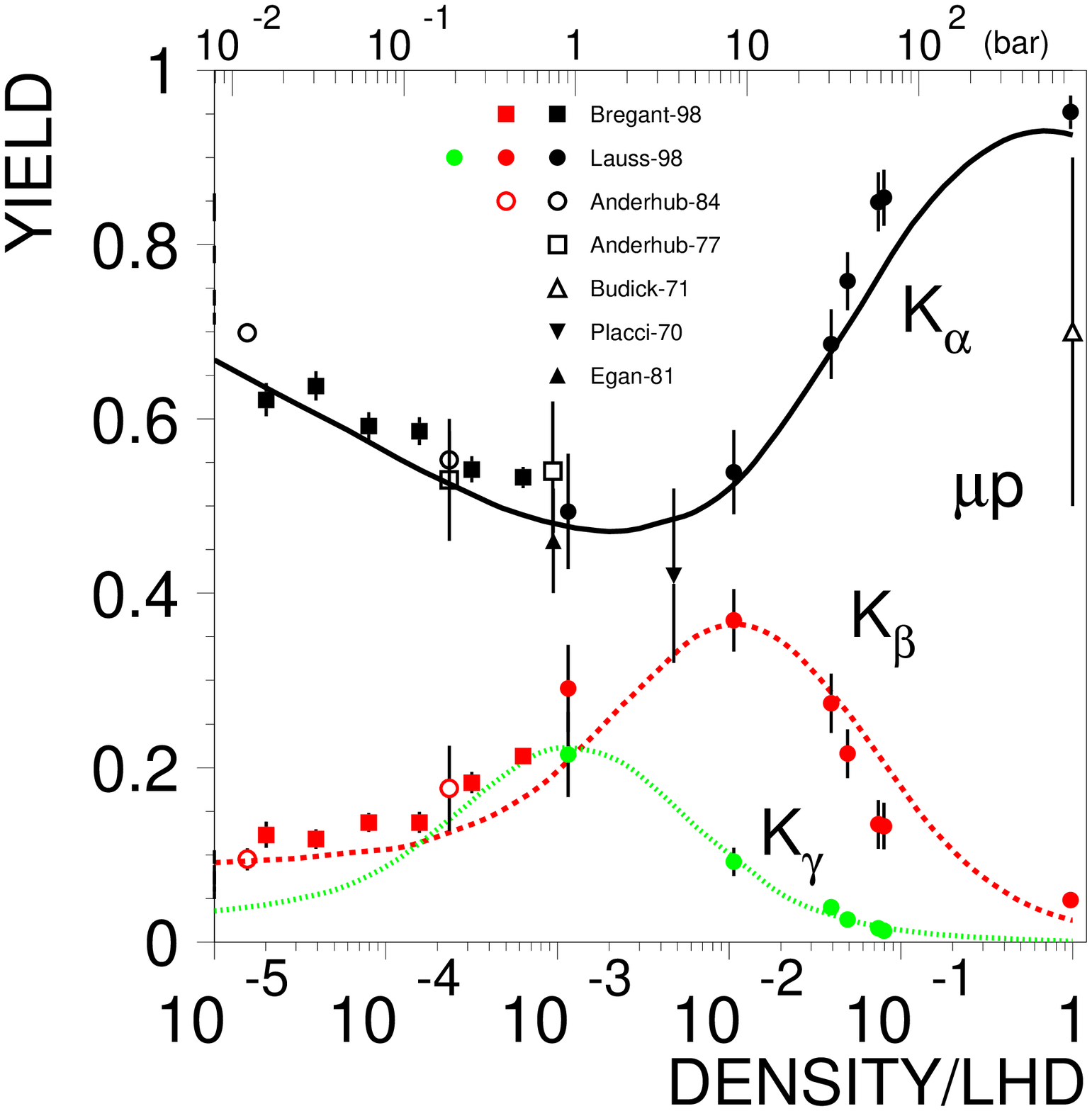}}%
\mbox{\epsfysize=65mm\epsffile{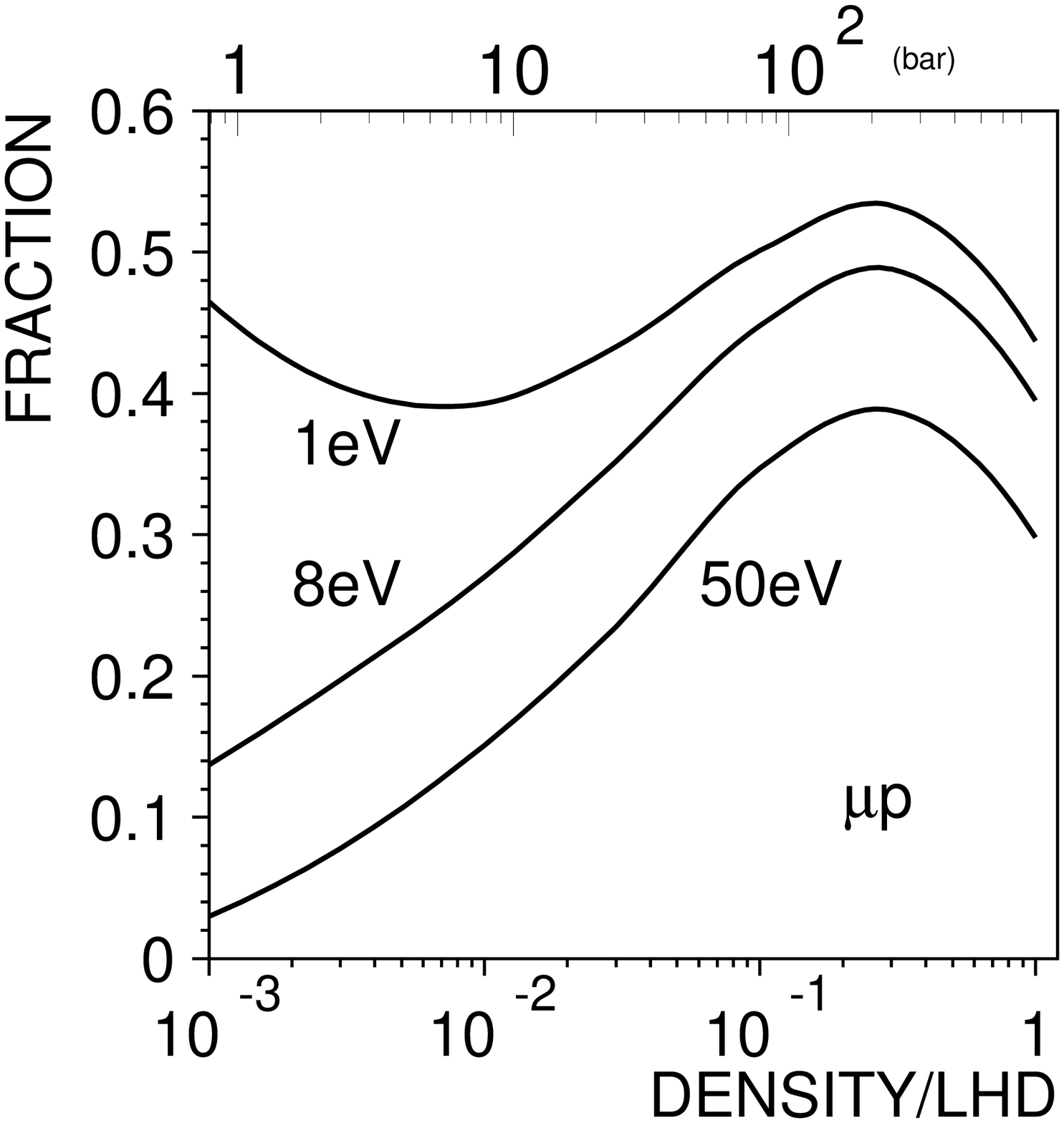}}
}
\caption{\label{FigXRayHE}%
(a) The density dependence of the $K_{\alpha}$, $K_{\beta}$, and
$K_{\gamma}$ yields in muonic hydrogen.
The theoretical curves are from \protect\cite{Ma94},
the experimental data from
\protect\cite{Pl70,Bu71,An77,Eg81,An81,An85,La98,Br98}.
(b) The density dependence of the high-energy components
($E\geq 1,\ 8,\ 50\;$eV)
in the $\mu p$ ground state after the atomic cascade calculated
in the model \cite{Ma94} with the Coulomb rates scaled by a factor
$k_C=0.5$ \cite{As95}.
}
\end{figure}

   Since the rates of the radiative transitions are well known, the
competition between the radiative and collisional processes can be used
for testing the collisional de-excitation rates by measuring the X-ray
yields at various densities.  The most suitable system for this study is
the muonic hydrogen where the X-ray yields are not suppressed by
absorption during the cascade.  Another convenient factor is that the
rates of the Auger de-excitation, which is the main collisional process,
have a weak energy dependence, therefore they are not strongly affected
by uncertainties in the kinetic energy distribution.
   The main features of the density dependence of the yields of the
$K$-lines were already fairly well explained by the MCM
\cite{BL80,Ma81}.  Figure~\ref{FigXRayHE}a shows the experimental data
in comparison with the recent calculations \cite{Ma94} that include, in
addition to the Auger de-excitation, the Coulomb transitions.

   The relative role of the collisional processes can be illustrated by
the following simplified calculation of the density dependence of the
$K_{\alpha}/K_{\beta}$ ratio using the method suggested in \cite{Ma84}.
The yields $Y_{K_{\alpha}}$ and $Y_{K_{\beta}}$ are given by the balance
equations:
\begin{eqnarray}
 Y_{K_{\alpha}} & \approx & p_{2} \; \approx \; p_{3} \;
    \frac{ \lambda^{\gamma}_{3\to 2}+
      \phi(\lambda^{Auger}_{3\to 2}+\lambda^{Coulomb}_{3\to 2})}
         {\lambda^{tot}_3} \label{YKa}
\\
 Y_{K_{\beta}} & = & p_{3} \;
   \frac{\lambda^{\gamma}_{3\to 1}}{\lambda^{tot}_3}
\ , \quad
 \lambda^{tot}_3=\lambda^{\gamma}_{3\to 2}+\lambda^{\gamma}_{3\to 1}
    +\phi(\lambda^{Auger}_{3\to 2}+\lambda^{Coulomb}_{3\to 2})
\label{YKb}
\end{eqnarray}
where $p_n$ are the populations of the atomic states $n=2,3$,
$\lambda^{\gamma}_{3\to 1}=\lambda^{\gamma}_{3P\to 1S}/3$ and
$\lambda^{\gamma}_{3\to 2}=\sum_l(2l+1)\lambda^{\gamma}_{3l\to 2P}/9$
are the effective rates of the radiative transitions,
$\phi$ is the hydrogen density in units of liquid hydrogen density
$N_0=4.3\cdot 10^{22}\mbox{\rm cm}^{-3}$ (LHD),
$\lambda^{Auger}_{3\to 2}$ and $\lambda^{Coulomb}_{3\to 2}$ are the
rates of the Auger and Coulomb transitions normalized to LHD.
Equations (\ref{YKa},\ref{YKb}) are valid at $\phi>0.05$ and
give the following dependence of the ratio
$Y_{K_{\alpha}}/Y_{K_{\beta}}$ on the density:
\begin{eqnarray}
 \frac{Y_{K_{\alpha}}}{Y_{K_{\beta}}} & = &
   \frac{\lambda^{\gamma}_{3\to 2}}{\lambda^{\gamma}_{3\to 1}} +
   \frac{\phi(\lambda^{Auger}_{3\to 2}
             +\lambda^{Coulomb}_{3\to 2})}{\lambda^{\gamma}_{3\to 1}}
   = 0.79 + 14.6 \phi \left( 1+
   \frac{\lambda^{Coulomb}_{3\to 2}}
        {\lambda^{Auger}_{3\to 2}} \right)  \label{YaYb}
\end{eqnarray}
If one neglects the Coulomb de-excitation, then Eq.~(\ref{YaYb})
gives $(Y_{K_{\alpha}}/Y_{K_{\beta}})^{th}=15$ at $\phi=1$
which is slightly below the corresponding experimental ratio
$(Y_{K_{\alpha}}/Y_{K_{\beta}})^{exp}=19.9\pm 2.5$ \cite{La98}.
At $\phi=0.08$ the difference is larger:
$(Y_{K_{\alpha}}/Y_{K_{\beta}})^{th}=2.0$
and $(Y_{K_{\alpha}}/Y_{K_{\beta}})^{exp}=6.4\pm 1.3$.
With the Coulomb transitions taken into account the theoretical ratio
gets closer to the experimental values as discussed in
\cite{La98,AM97}.
This observation can be considered as evidence that
some mechanisms in addition to the Auger effect, like the Coulomb
transitions, are needed for a better description of the X-ray yields.


\section{Kinetic Energy Distribution in Excited States}

   The evolution of the kinetic energy distribution during the atomic
cascade is very important because many collisional processes
are energy dependent (Table~1).
   In muonic atoms, it determines the initial energy distribution in the
ground state, which influences the diffusion of $\mu$-atoms
\cite{Ab97,Ha96,Ko98} and muon catalyzed fusion \cite{MCFPo,MCFCo}.
   A large fraction of the atoms is not thermalized\footnote{The atoms
with the kinetic energy much larger than the temperature are
called epithermal and the ones with $E\gg 1\;$eV --- 'hot' or
'highly-energetic'.} during the atomic cascade,
as it follows from the cascade calculation demonstrated in
Fig.\ref{FigXRayHE}b.

\begin{figure}
\mbox{\hspace{20mm} (a)\hspace{40mm} (b)\hspace{40mm}(c)}
\\[-0.5\baselineskip]
\mbox{%
\mbox{\epsfxsize=45mm\epsffile{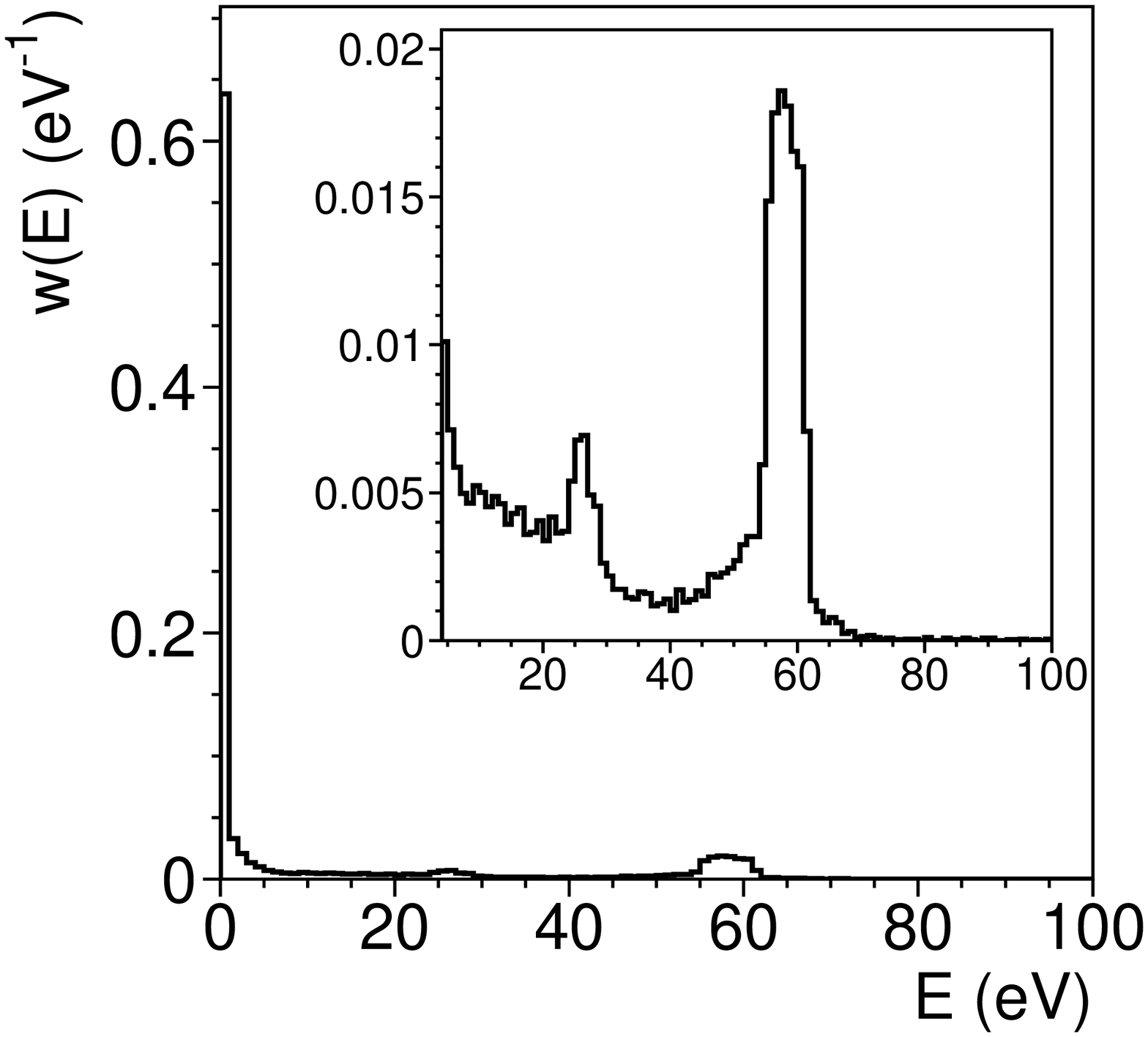}}%
\mbox{\epsfxsize=45mm\epsffile{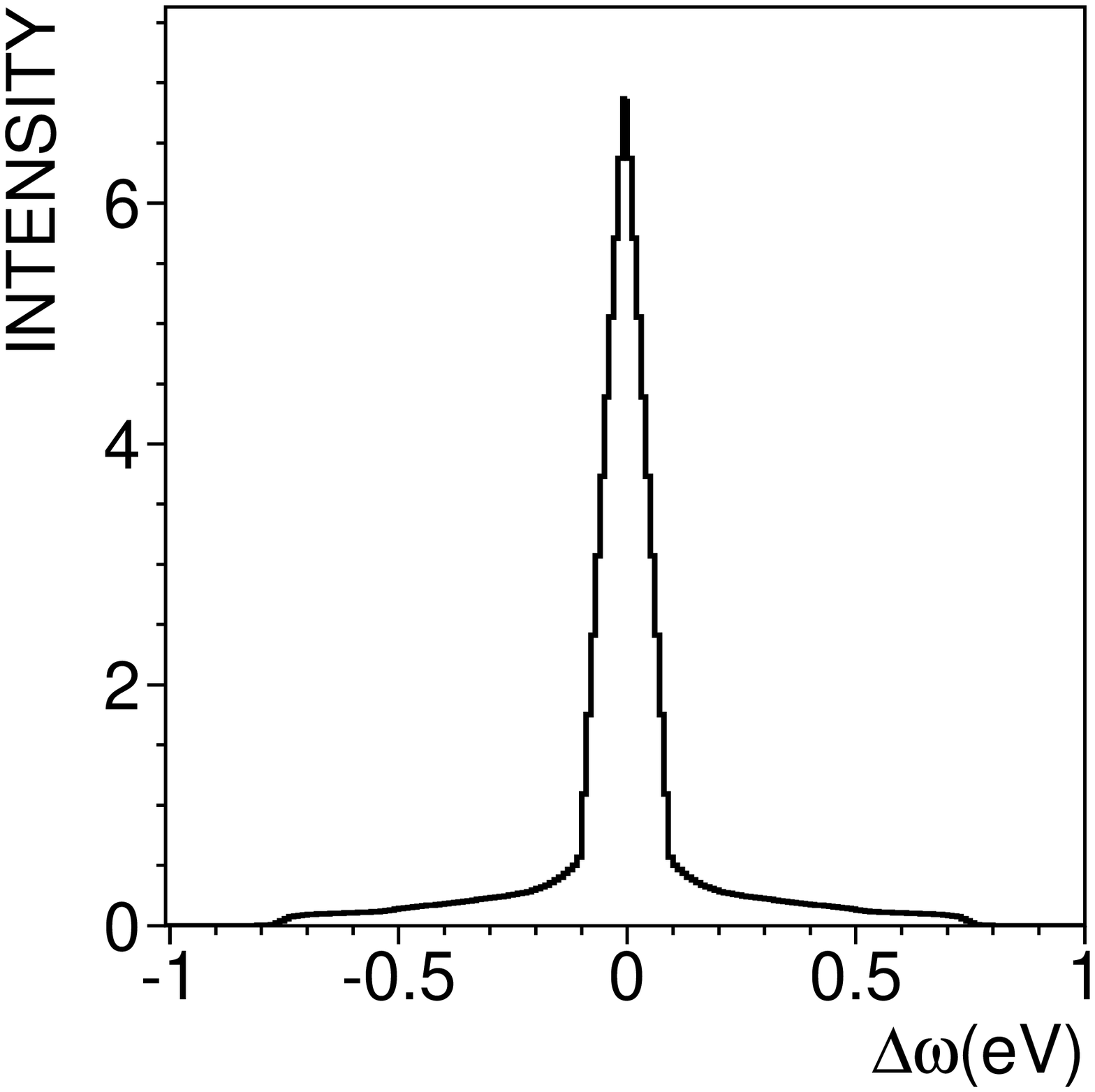}}%
\mbox{\epsfxsize=45mm\epsffile{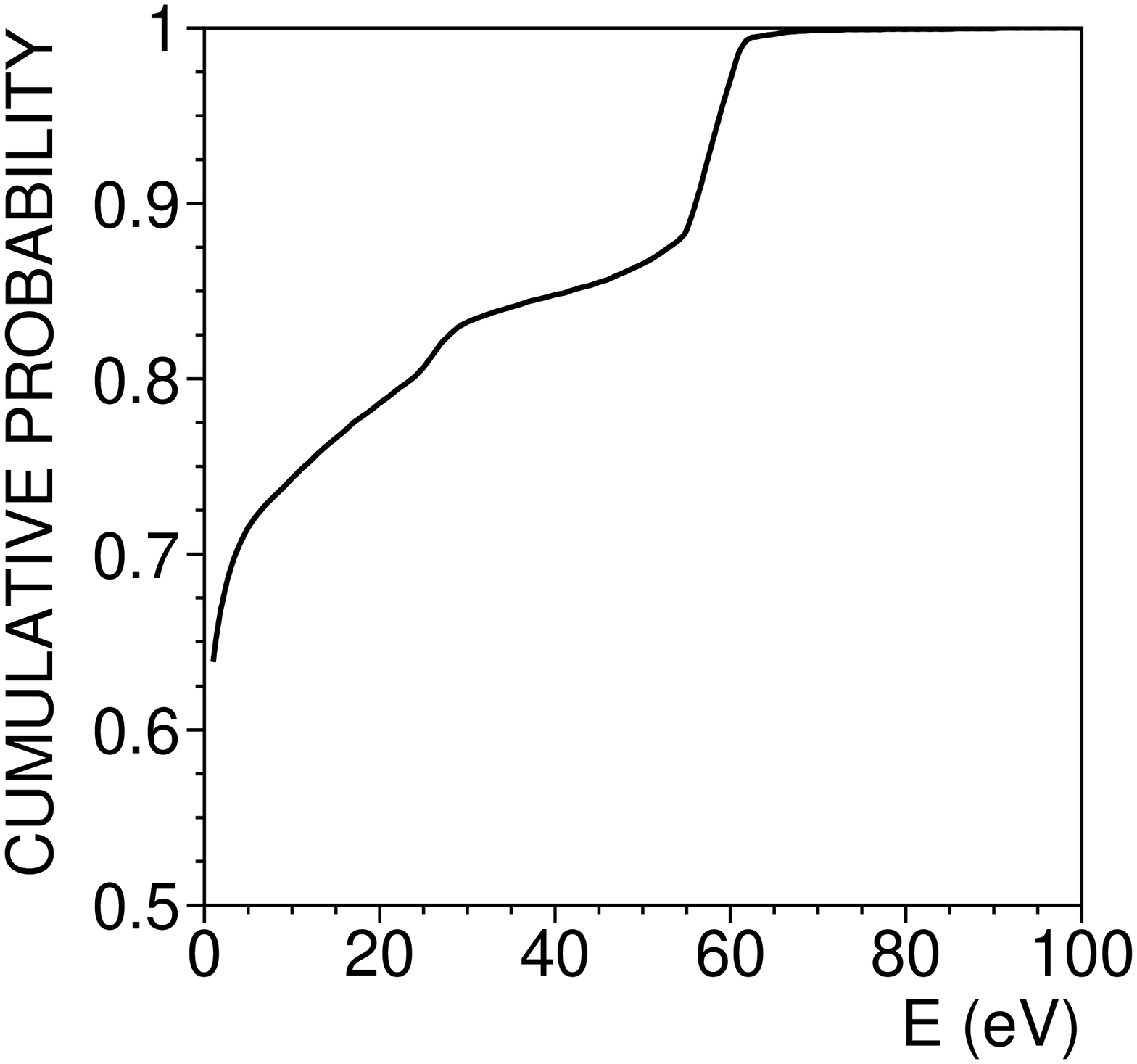}}
}
\caption{\label{FigmupDoppler}
(a) The kinetic energy distribution $w(E)$ at the instant of
the $3P\to 1S$ transition in muonic hydrogen at 15~bar,
(b) the corresponding Doppler broadening of the $K_{\beta}$ line and
(c) the cumulative energy distribution $W(E)$.
}
\end{figure}

   The kinetic energy distribution in the atomic cascade can be studied
with different methods.  Direct probes which are model independent are
based on the measurements of the Doppler broadening of the X-ray lines
in $\mu p$ and $\pi^-p$ and the $n$-TOF spectra in the reaction
$\pi^-p\to\pi^0+n$.
   Given the kinetic energy distribution $w(E)$ at the instant of the
radiative transition, the Doppler broadening of the X-ray line
$g(\Delta\omega)$ has the form:
\begin{eqnarray}
  g(\Delta \omega)  & = & \frac{1}{2\xi}
  \int_{(\Delta \omega/\xi)^2}^{\infty}        
  \frac{w(E)}{\sqrt{E}} dE
\ , \quad
  \xi  = \frac{\omega_0}{c} \sqrt{\frac{2}{M}} \label{DbwE}
\end{eqnarray}
where $\omega_0$ is the X-ray energy in the atomic rest frame
and $M$ is the mass of the atom.
A straightforward inversion of the transformation (\ref{DbwE})
gives the cumulative energy distribution $W(E)$:
\begin{eqnarray}
  W(E) \stackrel{def}{=} \int_{0}^{E} w(E') dE'  & = &
  2 \int_{_{0}}^{^{\xi\sqrt{E}}} g(x) dx -
  2 \xi\sqrt{E} g(\xi\sqrt{E}) \quad .  \label{WEDb}
\end{eqnarray}
Figure~\ref{FigmupDoppler} shows an example of the calculated
distributions for the $K_{\beta}$-transition in $\mu p$ atoms at 15~bar
which features characteristic peaks in the high-energy component
resulting from the Coulomb de-excitation discussed in
Sec.~\ref{SecCoulomb}.
   The Doppler broadening of the neutron TOF spectra from the reaction
$(\pi^-p)\to\pi^0+n$ is related to the kinetic energy distribution in a
similar way (see \cite{Cr91,Ba97,Sch98}).
   A good knowledge of the kinetic energy distribution is essential for
precise measurements of the $(\pi^-p)_{1S}$ nuclear width \cite{Si98}.
For example, the kinetic energy $T=0.5\;$eV corresponds to a Doppler
broadening of the $K_{\beta}$ line $\delta\Gamma=0.1\;$eV which is about
10\% of the nuclear width $\Gamma_{1S}=1\;$eV.

   Indirect methods of probing the kinetic energy distribution rely on
models of the kinetics.  In particular, the first evidence for
epithermal muonic atoms was found in muon catalyzed fusion (see
\cite{MCFCo} and references therein).
   Important results on the initial kinetic energy distribution in the
ground state were obtained from the diffusion of $\mu p$ and $\mu d$
atoms at low density \cite{Ab97,Ha96,Ko98}; they allow one to determine
the energy distribution in excited states using cascade models as
discussed in \cite{Ma94,Ko98}.
   Indirect methods exploiting energy dependent processes, like the muon
transfer in excited states \cite{Cz94T,La96,La98a}, were used for the
estimates of the average energy in excited states, however, they are not
sensitive to the details of the energy distributions.

\section{Elastic Collisions}


   The elastic collisions, where the principal quantum number of the
exotic atom is not changed, play two important roles in the atomic
cascade.  First, the Stark mixing is nothing but the elastic scattering
as long as only the rates of transitions between the $l$-sublevels of
the same $n$ are concerned.  Second, the energy losses in elastic
collisions lead to the deceleration of the epithermal exotic atoms.

The basic features of the elastic scattering can be explained in the
semiclassical approximation \cite{MP84}. The motion of
the $\mu p$ atom with the parabolic quantum numbers $(n_1,n_2,m)$
in the electric field of a hydrogen atom is described by the
effective potential\footnote{The quantization axis is
along the electric field, $n=n_1+n_2+|m|+1$.}
\begin{eqnarray}
     V(R) & = & \frac{g}{R^2} \zeta(R/a)
\ , \quad
       g   =   \frac{3n\Delta}{2m_{\mu p}}
\end{eqnarray}
where $\Delta=(n_1-n_2)$, $m_{\mu p}$ is the $\mu p$ reduced mass,
$\zeta(r)=(1+2r+2r^2){\rm e}^{-2r}$ is the electron screening factor,
$a$ is the electron
Bohr radius.
   Neglecting the electron screening, this corresponds to the well known
$R^{-2}$ potential with a large dimensionless coupling constant
$2Mg=\frac{3}{2}n\Delta M/m_{\mu p}\sim Mn^2/m_{\mu p}\gg 1$
($M$ is the reduced mass of the $\mu p+p$ system).
The $R^{-2}$ behaviour leads to the $E^{-1}$ dependence of the
elastic cross section.  In the limit $2Mg\gg 1$,  many partial waves
contribute to the scattering, and the differential cross section can
be approximated by the classical result ($g>0$):
\begin{eqnarray}
    \frac{d\sigma(E,\theta)}{d\Omega} & = &
        \frac{\pi g}{E} \; w(\theta) \label{XelCl} \\
    w(\theta) & = &
    \frac{\pi(\pi-\theta)}{(2\pi-\theta)^2\; \theta^2 \sin{\theta} }
    \approx \frac{1}{32 \sin^3{\theta/2}} \label{XelClw}
\end{eqnarray}
The average coupling constant for given $n$ is
$g=(n^2-1)/2m_{\mu p}$.
   The cross section (\ref{XelCl},\ref{XelClw}) has an unphysical
singularity at $\theta\to 0$ which must be regularized by taking the
electron screening into account.



   A semiclassical model of the Stark mixing was developed in
\cite{LB62} and used with some variations and refinements
in \cite{Ve,BL80,Ma81,TH97}. In this model, the exotic atom
moves along a straight line trajectory with a
constant velocity through the electric field of a hydrogen atom, and
the Stark transitions induced by this electric field are calculated.
Instead of straight line trajectories the classical trajectories
can be used as in \cite{Re88}.
   The characteristic scale of the Stark mixing cross section is
determined by the electron screening of the proton electric field, i.e.
by the size of the hydrogen atom which makes Stark mixing
the fastest collision process in the atomic cascade, see
Figs.~\ref{Figmuppip}.


   Recently the problem of the elastic scattering of exotic atoms in
excited states was treated in a quantum mechanical framework using the
adiabatic expansion \cite{PP96,PP98,PPG98}, the calculation being done
in a single-channel adiabatic approximation.  The results show
fair agreement with the semiclassical calculations
\cite{By96} for the collision energies $E>1\;$eV.  A detailed discussion
of the Stark collisions, including the relative role of transitions with
different $\Delta l=l_f-l_i$ can be found in \cite{PP98}.  It would be
desirable to extend the quantum mechanical model of the elastic
collisions by including the effects of the shift and width of the
$nS$ states as well as the coupling between different adiabatic terms.


   The elastic scattering leads to the deceleration of exotic atoms
if the kinetic energy is larger than the target temperature.
The effective deceleration rate for the $\mu p+p$ collisions is defined
by the transport cross section:
\begin{eqnarray}
   \lambda_n^{dec}(E) & = &
 N_0\; v\; \frac{2\;M_{\mu p}\;M_H}{(M_{\mu p}+M_H)^2}\;\sigma_n^{tr}(E)
, \\
   \sigma_n^{tr}(E) & = & \int (1-\cos\theta) d\sigma_n(E,\theta)
\end{eqnarray}
where $v$ is the atomic velocity, $M_{\mu p}$ and $M_H$ are the $\mu p$
and hydrogen masses correspondingly.
   Using the classical result (\ref{XelCl},\ref{XelClw}) one gets the
following estimation\footnote{Note that the transport cross section is
finite even without the electron screening.}
of the transport cross section:
\begin{eqnarray}
   \sigma_n^{tr}(E) & \approx &
   \frac{\pi^2(n^2-1)}{4m_{\mu p}E} \quad .
\end{eqnarray}
   The deceleration rate (Fig.~\ref{Figmuppip}) is a crude measure of
the slowing down; in general one uses the differential cross sections in
detailed kinetics calculations.
   The first estimation of the deceleration rates was done in
\cite{MP86}, semiclassical calculations were done in \cite{Cz96,By96},
and quantum mechanical calculations in \cite{PP96,PP98,PPG98}.

\section{Coulomb De-excitation}
\label{SecCoulomb}

   The Coulomb de-excitation (Table~1), where the transition energy is
shared between the colliding particles, is an important acceleration
mechanism producing 'hot' ($E\gg 1\;$eV) exotic atoms (see
Fig.~\ref{FigmupDoppler}a).
   The highly-energetic $\pi^-p$ atoms were discovered experimentally in
\cite{Cz63}, and a multicomponent structure of the energy distribution
was found in \cite{Cr91}.
Recent experiments in liquid and gaseous hydrogen
\cite{Ba97,Sch98} found the shape of the kinetic energy distribution to
be consistent with the Coulomb mechanism.
   The cascade calculations \cite{As95} show that the $\pi^-p$ atoms
with kinetic energy $E\geq 50\;$eV are not significantly decelerated
between the Coulomb de-excitation and the nuclear reaction. For the
atoms with $E\leq 20\;$eV the deceleration is important and the Coulomb
peaks are expected to be smeared out.

   The earlier calculations of the Coulomb transitions
\cite{BF,Me88,Cz90,Cz94CT} were rather controversial.  The latest
calculations \cite{PS96,So98,Kr98} seem to clarify the theoretical
picture.
   This removes one of the uncertainties in the cascade calculation of
the past where a Coulomb rate-tuning parameter was used to normalize the
calculated high-energy component to the measurements in pionic hydrogen
\cite{As95}.
   It is not excluded that the Coulomb de-excitation can be part of a
multistep process.  For instance, the formation of excited molecular
states can be followed by a Coulomb-like decay \cite{Ma90,Taq,Fr95}.

\section{Atomic Cascade in H-D Mixtures}

\begin{figure}
\mbox{\epsfysize=60mm\epsffile{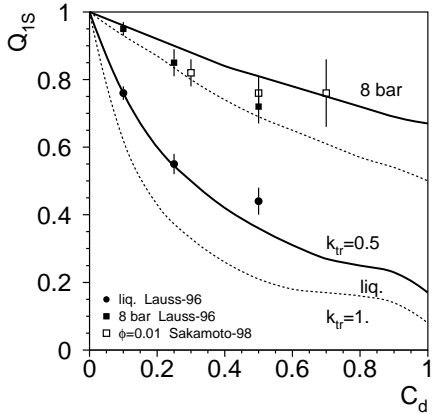}}\vspace{-3mm}
\caption{\label{FigQ1S}
The $Q_{1S}$ factor for the $\mu p$ cascade in HD mixture vs.
deuterium fraction $C_d$.
The data are from \cite{La96,Sa98}; the theoretical curves
\cite{AM97} were calculated with the transfer rates \cite{GPS93}
scaled by the factor $k_{tr}=0.5$ (solid lines) and
$k_{tr}=1$ (dashed lines).
}
\end{figure}

   The atomic cascade in mixtures of hydrogen isotopes provides
additional
information on the collisional processes in excited states.  The muon
transfer in excited states was studied in detail in the recent
experiments \cite{La96,La98a,Sa98} by measuring the relative yields of
the $\mu p$ and $\mu d$ $K$-lines in liquid and gaseous HD mixtures.
   Figure~\ref{FigQ1S} shows a comparison of the experimental data with
the cascade calculations for the probability $Q_{1S}$ that
the muon captured initially by the hydrogen in a HD mixture reaches the
$\mu p$ ground state (the fraction $(1-Q_{1S})$ is transferred to
$\mu d$ during the atomic cascade).
   Since the rates of the muon transfer
$(\mu p)_n+d\leftrightarrow(\mu d)_n+p$ are strongly energy dependent
\cite{MP84,Kr88,GPS93,Cz94T}, the $Q_{1S}$ factor is very sensitive to
the kinetic energy distribution in the excited states
\cite{Ma90,Kr88,Cz94T,Cz90Casc,AM97}.

   The theoretical models tend to predict a stronger dependence of
$Q_{1S}$ on the deuterium fraction and the density than experimentally
observed 
(this so-called "$Q_{1S}$-problem" is known since the first experimental
results were obtained from the kinetic analysis of muon catalyzed fusion
\cite{MCFPo,MCFCo}).  The agreement between theory and experiment
improves if the transfer rates are scaled by a factor of about 0.5
\cite{AM97}.  This suggests that some mechanism may still be missing in
the cascade model.
   One candidate is the resonant molecular formation in excited states
(resonance side-path mechanism \cite{Fr95}) which can produce a
significant inverse transfer $\mu d\to\mu p$ leading to an enhancement
of $Q_{1S}$ as discussed in \cite{Wa98}.

\section{Conclusion}

   The current cascade models are advancing from a phenomenological
level towards straightforward detailed calculations of the kinetics.
The recent progress in quantum mechanical calculations of the scattering
of exotic atoms in excited states makes it possible to get rid of tuning
parameters and ad hoc assumptions in future calculations.  The ultimate
goal of {\it ab initio} cascade calculations does not seem to be
unrealistic any more.

   A detailed knowledge of atomic cascade is essential for
precision X-ray spectroscopy of the $\pi^-p$ atom where the
Doppler-broadening corrections must be taken into account in order to
measure the nuclear width of the $1S$ state.  The new generation of
precise experiments with the $\mu p$ and $\pi^-p$ atoms \cite{Si98}
will make it possible to study the kinetics of the atomic cascade in
great detail and to perform critical tests of the theoretical methods used
in the calculations of various reactions with excited exotic atoms.
The results of theses studies will also be very useful for other
systems like $K^-p$ and $\bar{p}p$.

\section*{Acknowledgements}

   The author thanks A.~Badertscher, M.~Daum, P.~Froelich,
P.F.A.~Goudsmit, F.~Kottmann, B.~Lauss, H.J.~Leisi, L.I.~Ponomarev,
V.P.~Popov, J.~Schottm\"{u}ller, L.M.~Simons, E.A.~Solov'ev, D.~Taqqu,
and J.~Wallenius for fruitful discussions.



\end{document}